\documentclass[11pt]{article}
\usepackage{hyperref}
\usepackage{amsmath}
\usepackage{amssymb}
\renewcommand{\title}[1]{%
    \bigskip%
    \begin{center}%
    \Large\bf #1%
    \end{center}%
    \vskip .2in}

\renewcommand{\author}[1]{%
    {\begin{center}
    #1
    \end{center}}}
\newcommand{\address}[1]{\vspace{-1.7em}\vspace{0pt}
    {\begin{center}
    \it #1
    \end{center}}}

\begin{document}


\title{A new formulation of non-relativistic diffeomorphism invariance}

\author
{
Rabin Banerjee  $\,^{\rm a,b}$,
Arpita Mitra    $\,^{\rm a, e}$,
Pradip Mukherjee $\,^{\rm c,d}$}
\address{$^{\rm a}$S. N. Bose National Centre 
for Basic Sciences, JD Block, Sector III, Salt Lake City, Kolkata -700 098, India }

\address{$^{\rm c}$Department of Physics, Barasat Government College,\\Barasat, West Bengal

 }

\address{$^{\rm b}$\tt rabin@bose.res.in}
\address{$^{\rm d}$\tt mukhpradip@gmail.com}
\address{$^{\rm e}$\tt arpita12t@bose.res.in}

\begin{abstract} 
We provide a new formulation of nonrelativistic diffeomorphism invariance. It is generated by localising the usual global Galilean Symmetry. The correspondence with the type of diffeomorphism invariant models currently in vogue in the theory of fractional quantum Hall effect has been discussed. Our construction is shown to open up a general approach of model building in theoretical condensed matter physics. Also, this formulation has the capacity of obtaining Newton - Cartan geometry from the gauge procedure.
\end{abstract}

\section{Introduction}
Recently there has been a spate of papers in the literature \cite{SW, H, SON, AHH} which use the non relativistic diffeomorphism invariance to analyse the motion of two dimensional trapped electrons which is directly connected with the theory of Fractional Quantum Hall Effect (FQHE). The relevant field theories involve some variant of the Schrodinger field theory in the framework of 3-d manifold with time running universally. These works are done at the effective field theory level and no attempt
has been made to formulate the theories from  some fundamental premises. 
The theories thus formulated are also endowed with gauge symmetry that correspond to the trapping fields. Thus in these theories gauge invariance and nonrelativistic diffeomorphism  invariances are introduced separately. It would be indeed very nice if these symmetries emerge systematically from some basic principles.

   The way the Galilean symmetric model of the Schrodinger field is conventionally attributed diffeomorphism invariance to eventually discuss FQHE also raises some more questions. The set of transformation of the basic fields is not only model specific but also different for different types of diffeomorphism parameters. The basic fields obey tensor transformation rules when the diffeomorphism invariance is given by $x^i \to x^i + \xi^i$ with time independent $\xi$. When attempts are made to include time varying diffeomorphism parameters special transformation rules of the fields are assumed which lack motivation. In fact the authors of \cite{SW} have themselves admitted that they have found the transformations by `trial and error'.

 We further observe that an important requirement of such a construction would be to obtain the global Galilean group of symmetry in the flat limit. In the usual approaches \cite{SW} it is not apparent how the translation is distinguished from spatial rotation in this limit. Secondly, an intriguing situation occurs for the case of Galilean boost where to restore the Galilean symmetry in the flat limit one has to invoke a relation between the gauge parameter and the boost parameter \cite{SW}! Obviously it is very difficult to motivate such an interrelation.

           The last issue in particular shows that it may not always be consistent to impose diffeomorphism in the 3-d space by invoking effective field theories. One requires a systematic approach towards diffeomorphism in the 3-d space. In this connection we note that in the nonrelativistic scenario there is no natural concept of spacetime manifold. The geometric approach developed by Cartan \cite{Cartan} and later elaborated by other stalwarts \cite {Havas, ANDE, TrautA, Kunz, Kuch, Daut, EHL, MALA} is of no direct avail in formulating diffeomorphism in the 3-d space. This has been discussed in several recent papers \cite{PLP, BP}.
           
           In this paper we present a systematic procedure to obtain nonrelativistic diffeomorphism invariance. The method is to localize the Galilean symmetry of a nonrelativistic field theory. The starting point is a nonrelativistic field theory invariant under global Galilean transformations with constant values of the translation, rotation and boost parameters. We then make the Galilean transformations local. Following the nature of Galilean spacetime the time translation parameter is assumed to be function of time whereas the other parameters are assumed to be functions of space and time. Anticipating the emergence of nonrelativistic diffeomorphism invariance in space we construct a set of coordinate axis at every spatial point which are trivially connected with the global coordinate system. This connection is subsequently shown to be non trivial. As an effect of making the parameters local the transformations of the derivatives of the fields are now different from that under global transformation. The Galilean invariance of the original model is thus broken. To restore the symmetry it is required to introduce new fields whose variations would cancel this difference. These new fields enter through covariant derivatives that replace the ordinary derivatives. The covariant derivatives are defined in two stages. In the first stage the covariant derivatives with respect to the coordinate basis are defined. New fields  $A_t$ and $A_k$ (k=1,2,3) are introduced at this stage. Finally additional fields are introduced to define covariant derivatives in the local basis. Additional fields $\theta(t), \Psi^k(t,r), {\Sigma_a}^{k}(t,r)$ appear now. The transformation rules of the new fields are chosen so as to transform the local covariant derivatives in the same way as the transformations of the ordinary derivatives under global transformation. Apart from restoring Galilean invariance this localisation, subject to a further restriction,   introduces diffeomorphism in 3-d space as we shall see. 

\section{Localisation of Galilean symmetry}
Let us consider a non relativistic model invariant under global Galilean transformations given by the action
\begin{equation}
S = \int dt d^3 x {\cal{L}}\left(\phi,\partial_t{\phi}, \partial_k{\phi}\right)
\label{action}
\end{equation}
Our strategy will be to try to make the symmetry of the model (\ref{action}) local. We  assume that $\phi$ is a scalar under the Galilean transformations. Apart from  this simplification our model is perfectly general. The global Galilean transformations under which (\ref{action}) is invariant are given by
\begin{equation}
\xi^{0}=-\epsilon,~~~~~~\xi^{i}=\epsilon^{i}+{C^{i}}_{jk}\omega^{j}x^{k}-v^{i}t \label{globalgalilean}
\end{equation}
where $\epsilon$, $\epsilon^{i}$, $\omega^{i}$ and $v^{i}$ are constants and ${C^{i}}_{jk}$ are the structure constants of the rotation group. These structure constants are antisymmetric under interchange of any pair of indices.

Localisation of (\ref{globalgalilean}) demands careful consideration. Nonrelativistic Newtonian gravity was formulated on 4-dimensional spacetime manifold by Elie Cartan. There the gravitational structure was defined in terms of an affine connection compatible with the temporal flow $t_{\mu}$ and a rank-three spatial metric $h^{\mu\nu}$. Keeping this in view we localise (\ref{globalgalilean}) in the following way
\begin{equation}
\xi^{0}=-\epsilon\left(t\right),~~~~~~\xi^{i}=\epsilon^{i}\left(t, {\bf{r}}\right)+{C^{i}}_{jk}\omega^{j}\left(t, 
{\bf{r}}\right)x^{k}-v^{i}\left(t, {\bf{r}}\right)t \label{localgalilean}
\end{equation}
For later calculations it will be advantageous to write $\xi^i$ as,
\begin{equation}
\xi^i = \eta^i\left(t, {\bf{r}}\right) - v^i\left(t, {\bf{r}}\right)t
\end{equation}
where $\eta^i\left(t, {\bf{r}}\right) = \epsilon^{i}\left(t, {\bf{r}}\right)+{C^{i}}_{jk}\omega^{j}\left(t, 
{\bf{r}}\right)x^{k}$. 

         To achieve the localisation, it will be helpful to understand the mechanism of global Galilean invariance under (\ref{globalgalilean}). Under the general coordinate transformation $x^\mu \to x^\mu + \xi^\mu$ the action (\ref{action}) changes by
\begin{equation}
\Delta S = \int dt d^3x \Delta{{\cal{L}}}
\end{equation}
with
\begin{equation}
\Delta {{\cal{L}}} = \delta_0{{\cal{L}}} + \xi^{\mu}\partial_{\mu}{{\cal{L}}}+ \partial_{\mu}\xi^{\mu}{{\cal{L}}}\label{formvariation}
\end{equation}
 Here $\delta_0$ denotes the form variation given by 
\begin{equation}
\delta_0 \psi = \psi^{\prime}\left({\bf{r}}, t\right) - \psi\left({\bf{r}}, t\right)
\end{equation}
and $x^{\mu} \equiv t,x,y,z$. For invariance we require {\footnote{Of course invariance is preserved if ${\cal{L}}$ changes by a total derivative. However we only consider such theories where 
${\Delta{\cal{L}}} = 0$.}}
\begin{equation}
\Delta {{\cal{L}}} =\delta_0{{\cal{L}}} + \xi^{\mu}\partial_{\mu}{{\cal{L}}}+ \partial_{\mu}\xi^{\mu}{{\cal{L}}}= 0.
\label{22}
\end{equation}
For global Galilean transformation this is ensured by two conditions, 
\begin{enumerate}
\item    $\partial_{\mu}\xi^{\mu} = 0$, which may be explicitly checked using (\ref{globalgalilean}). So the invariance condition (\ref{22}) reduces to 
\begin{equation}
\Delta {{\cal{L}}} = \delta_0{{\cal{L}}} + \xi^{\mu}\partial_{\mu}{{\cal{L}}} = 0\label{reduced}
\end{equation}
\item The form variations of the field and its derivatives are given by 
\begin{eqnarray}
&\delta_0\phi = \epsilon\partial_t\phi - \epsilon^{k}\partial_k\phi - C^{i}_{jk}\omega^{j}x^{k}\partial_i\phi + tv^{k}\partial_k\phi - imv^{k}x_k \phi\nonumber\\
&\delta_0 \partial_k\phi=\epsilon\partial_{t}(\partial_{k}\phi)-\eta^{i}\partial_{i}(\partial_{k}\phi)+
\left(v^{i} t\partial_{i}-imv^ix_i\right)\partial_{k}\phi-{C^{l}}_{mk}\omega^{m}\partial_{l}\phi-imv_k\phi\nonumber\\
&\delta_0 \partial_t\phi=\epsilon\partial_{t}(\partial_{t}\phi)-\eta^{i}\partial_{i}(\partial_{t}\phi)+\left(v^{i} t\partial_{i}-imv^i x_i\right)\partial_{t}\phi+v^{k}\partial_{k}\phi
\label{delkphi}
\end{eqnarray}
In writing the form variation of $\phi$ we have exploited the fact that it is a complex Galilean scalar \cite{JP, MC}. In particular this explains the appearance of the last term in the transformation $\delta_0\phi$. The variations (\ref{delkphi}) are such that the simplified condition (\ref{reduced}) is satisfied.
\end{enumerate}
 
     When the symmetry group is local Galilean it will be advantageous to introduce local coordinates $x^a$ at each point of space. The corresponding local basis ${\bf{e}}^a$ is in general related with the global basis ${\bf{e}}^k$ by,
\begin{equation}
{\bf{e}}^a = {\Lambda^a}_k{\bf{e}}^k
\label{basis}
\end{equation}
where ${\Lambda^a}_k$ is the vielbein. Now, in flat space time it is always possible to choose ${\Lambda^a}_k = \delta_k^a$ assuming inertial coordinates. In that case
the vectors in the local basis are orthogonal just as happens for the global basis. 


The first step in the process of localisation is to convert the ordinary derivatives into covariant derivatives with respect to the global coordinates. To begin with, introduce the gauge fields $A_t$ and $A_k$ such that,
\begin{eqnarray}
D_k\phi=\partial_k\phi+iA_k\phi\nonumber\\
D_t\phi=\partial_t\phi+iA_t\phi \label{firstcov}
\end{eqnarray}
We retain the freedom of choosing the transformation rule of $A_t$ and $A_k$ at this stage.

As explained above the intermediate covariant derivatives (\ref{firstcov}) will now be used to define the covariant derivatives with respect to the local coordinates which will be required to have the appropriate transformation property to ensure the local Galilean symmetry. Introducing additional new fields $\theta (t)$, $\Psi^k(t, {\bf{r}})$ and 
${\Sigma_a}^k(t, {\bf{r}})$ we define the new covariant derivatives as,
\begin{eqnarray}
\nabla_a\phi={\Sigma_a}^{k} D_k\phi\nonumber\\
\nabla_t\phi=\theta(D_t \phi+\Psi^k D_k\phi)\label{finalcov}
\end{eqnarray}

The term involving $\Psi^k$ might appear mysterious. However the process of gauging naturally leads to a construction of the covariant derivative w.r.t the local time that involves both $D_t\phi$ and $D_k\phi$ type terms. Similar features prevail when the Poincare group is gauged \cite{Blagojevic:2002du}.

After a long algebra we find that these transform covariantly, 
provided the additional fields transform as
\begin{eqnarray}
&{\delta}_0 A_{k} = -{\partial}_k {\eta}^i A_i +t{\partial}_k v^iA_i + \epsilon \dot{A}_k - {\eta}^i {\partial}_iA_k +tv^i{\partial}_iA_k + mv_k +  m{\partial}_kv^ix_i-m{\Lambda^a}_k v_a\nonumber\\
&\delta_0A_t=m{\dot{v}}^ix_i+\epsilon{\dot{A}}_t+\dot{\epsilon}A_t-\eta^i
\partial_iA_t+v^i t\partial_i A_t-{\dot{\eta}}^iA_i+{\dot{v}}_i tA_i+v^i A_i +m{\Lambda^a}_k\Psi^k v_a
\label{delA}
\end{eqnarray}
and
\begin{eqnarray}
&\delta_0\theta=-\theta\dot{\epsilon}+\epsilon\dot{\theta}\nonumber\\
&\delta_0\Psi^k=\epsilon{\dot{\Psi}}^k+\dot{\epsilon}\Psi^k-\frac{1}{\theta}v^b{\Sigma_b}^{k}+\frac{\partial}{\partial t}({\eta}^k-tv^k)-(\eta^i-v^i t) \partial_i\Psi^k-t\Psi^i\partial_i v^k+\Psi^i\partial_i\eta^k\nonumber\\
&\delta_0 {\Sigma_a}^{k}=\epsilon{\dot{\Sigma}_a}^{k}+ {\Sigma_a}^{i}\partial_{i}\eta^{k}-t{\Sigma_a}^{i}\partial_{i}v^{k}+
{C^{b}}_{ca}\omega^{c}{\Sigma_b}^{k} - \eta^{i}\partial_{i}{\Sigma_a}^{k}+t v^{i} \partial_{i} {\Sigma_a}^{k}\label{delth}
\end{eqnarray}
where  ${\Lambda^a}_k$ is the inverse of ${\Sigma_a}^{k}$, 
\begin{equation}
{\Sigma_a}^{k}{\Lambda^a}_l=\delta^{k}_{l}
\label{sl}
\end{equation}
Observe that the transformations (\ref{delA}, \ref{delth}) are, in general, different from those given in \cite{H}. For the special case of $\xi^0=-\epsilon(t)=0$ and $\xi^i=\epsilon^i({\bf{r}})+{C^{i}}_{jk}\omega^{j}({\bf{r}})x^k$ (i.e. time independent translation and rotation), they agree, as will be shown subsequently (around (\ref{sw})). That the agreement does not hold in general is not surprising as our transformations involve fields $\Psi^k$ which are not there in \cite{H}. Only in the special case mentioned above $\Psi^k$ may be taken to be zero in which case the agreement holds.

It is easy to show that, 
\begin{equation}
\delta_0 {\Lambda^a}_{k}=\epsilon{\dot{\Lambda}^a}_{k}- {\Lambda^a}_{l}\partial_{k}\eta^{l}+t{\Lambda^a}_{l}\partial_{k}v^{l}-
C^{a}_{ec}\omega^e{\Lambda^c}_{k} - \eta^{i}\partial_{i}{\Lambda^a}_{k}+t v^{i} \partial_{i} {\Lambda^a}_{k}
\label{delLamb}
\end{equation}
We have finished the first stage of localisation of the Galilean symmetry (\ref{globalgalilean}). If in the Lagrangian 
${\cal{L}}\left(\phi, \partial_t{\phi}, \partial_k{\phi}\right)$ we substitute $\partial_t{\phi}$, $\partial_k{\phi}$ by  
 $\nabla_t{\phi}$, $\nabla_a{\phi}$ such that 
$$
{{\cal{L}}\left(\phi, \partial_t\phi, \partial_k\phi\right)} \to
 {{\cal{L}^{\prime}}\left(\phi, \nabla_t\phi, \nabla_k\phi\right)}
$$
then under the local Galilean transformations the modified Lagrangian ${{\cal{L^\prime}}}$ satisfies
\begin{equation}
\delta_0{{\cal{L^\prime}}} + \xi^{\mu}\partial_{\mu}{{\cal{L^\prime}}} = 0
\end{equation}
which is the analogue of the condition for global Galilean invariance given in (\ref{reduced}) where $\partial_\mu\xi^\mu = 0$.
This condition ($\partial_\mu\xi^\mu = 0$) does not however hold when the Galilean transformations are localised. Thus the modified Lagrangian ${{\cal{L^\prime}}}$ is still not invariant under local Galilean transformations.
We require a further modification
as,
\begin{equation}
{{\cal{L}}} = \Lambda{{\cal{L^\prime}}}
\end{equation}
To make the action invariant under the whole local transformation we have to satisfy (\ref{22}). This in turn leads to the following requirement
\begin{equation}
\delta_0\Lambda + \xi^\mu \partial_\mu\Lambda 
+ \partial_\mu \xi^\mu\Lambda=0
\label{lambda}.
\end{equation}
It is not difficult to find a $\Lambda$, which solves the above equation.
We make the following Ansatz
\begin{equation}
\Lambda = \frac{M}{\theta}
\end{equation}
where 
\begin{equation}
 M = det{\Lambda^a}_k\label{M}.
\end{equation}
Now we will show that this expression of $\Lambda$ satisfies (\ref{lambda}). The transformation of M can be calculated using the well known formula of differentiation of a determinant.
\begin{equation}
\delta_0 M=-M{\Lambda^a}_{k}\delta_0{\Sigma_a}^{k}
\label{delM}
\end{equation}
The transformation rule of  $\theta$ is obtained from (\ref{delth}) and $\delta_0 M$ can be calculated using $\delta_0{\Sigma_a}^{k}$ from (\ref{delth}) in (\ref{delM}). By direct substitution of $\delta_0\Lambda$ obtained in this way and $\xi^{\mu}$ from (\ref{localgalilean}) one can verify that equation (\ref{lambda}) holds. Note that, the transformation rule of $\Lambda$ is already predetermined by the transformation (\ref{delth}) of the new fields introduced in order to transform the covariant derivatives with respect to global coordinate to the corresponding covariant derivative with respect to  the local coordinates (see equation (\ref{firstcov})). Thus the satisfaction of (\ref{lambda}) is indeed remarkable. Specifically it points to the logical consistency of our construction.
 
   We thus derive the rules of localising the Galilean symmetry of a nonrelativistic model. The algorithm is as follows. Introduce local coordinates at each point of 3-d space. The local basis is trivially connected to the coordinate basis by (\ref{basis}). If the original theory is given by the action
\begin{equation}
S = \int dt d^3x {\cal{L}}\left(\phi, \partial_t \phi, \partial_k \phi\right)
\end{equation}
invariant under the global Galilean transformation
\begin{equation}
x^{\mu}\rightarrow x^{\mu}+\xi^{\mu}
\end{equation}
where $\xi^{\mu}$ is defined in (\ref{globalgalilean}), then
\begin{equation}
S = \int dt d^3x \frac{M}{\theta}{\cal{L}}\left(\phi, \nabla_t\phi, \nabla_a\phi\right)
\label{localaction}
\end{equation}
is invariant under the corresponding local Galilean transformations (\ref{localgalilean}). 

Observe that till this point we have worked solely on a flat space. Since we localised the global symmetry it was essential to introduce the local coordinates vis-a-vis the global ones. The two bases are related by (\ref{basis}). This procedure is completely analogous to what is done in the localisation of the Poincare symmetry \cite{Blagojevic1}.
\section{Nonrelativistic diffeomorphism invariance}
The importance of the construction (\ref{localaction}) cannot be overemphasized. As we now show this construction naturally leads to nonrelativistic 3-d diffeomorphism invariance. In order to achieve this we observe that $\xi^0=-\epsilon$ should be vanishing. Then the local Galilean transformation is equivalent to the transformation 
\begin{equation}
x^i\longrightarrow x^i+\xi^i\left({\bf{r}, t}\right)
\label{3ddiff}
 \end{equation}
where $\xi^i$ is an arbitrary function of ${\bf{r}}$ and $t$. \footnote{Note the connection between the local and the global basis (\ref{basis})}. From the first of the set of transformations (\ref{delth}) we find that when $\epsilon=0, \theta=$ constant. Without any loss of generality we can take $\theta= 1$. Now we introduce the `metric tensor' $g_{ij}$ 
\begin{equation}
g_{ij}=\delta_{cd}{\Lambda^c}_i {\Lambda^d}_j
\label{metric1}
\end{equation}
Using the transformation relation of ${\Lambda^a}_k$ from (\ref{delLamb}) we find 
\begin{equation}
\delta_0 g_{ij} =-\xi^k \partial_k g_{ij}-g_{ik}\partial_j\xi^k-g_{kj}\partial_i\xi^k
\label{diff} 
\end{equation}
From the transformation relation (\ref{diff}) it appears that $g_{ij}$ may serve as a viable definition of the metric in nonrelativistic diffeomorphism invariant 3-dimensional curved space. This is further supported from the observation that $M=\sqrt{g}$ where $g$ is the determinant of $g_{ij}$. This relation follows from the definitions of $M$ (\ref{M}) and $g_{ij}$ (\ref{metric1}). Using these in (\ref{localaction})  the local Galilean invariant action reduces  to  
\begin{equation}
S = \int dt d^3x \sqrt{g}{\cal{L}}\left(\phi, \nabla_ t\phi, \nabla_ a\phi\right)
\label{localaction1}
\end{equation}
It is easy to appreciate that this theory (\ref{localaction1}) is invariant under 3-d nonrelativistic diffeomorphism (\ref{3ddiff}).

      As a concrete example let us consider the Schrodinger field theory
\begin{equation}
S = \int dt  \int d^3x  \left[ \frac{i}{2}\left( \phi^{*}\partial_{t}\phi-\phi\partial_t\phi^{*}\right) -\frac{1}{2m}\partial_k\phi^{*}\partial_k\phi\right]
\label{globalaction} 
\end{equation}
which can be shown to be invariant under the global Galilean transformations (\ref{globalgalilean}) as follows.

 At first we have to compute the variation of $\mathcal{L}$ under (\ref{globalgalilean}) as follows,
\begin{align}
{\delta}_0 \mathcal{L} & = {\mathcal{L}}'-\mathcal{L}\nonumber\\& =\frac{i}{2}[({\delta}_0 {\phi}^{*}) {\partial}_{t}\phi+
{\phi}^{*}{\delta}_{0}({\partial}_{t}\phi)+\frac{i}{m}\delta_{0}(\partial_{k}\phi^{*})\partial_{k}\phi]+ c.c.
\label{c}
\end{align}
Analyzing the individual terms in (\ref{c}) we will get,
\begin{equation}
\frac{i}{2}(\delta_{0}\phi^{*})\partial_{t}\phi=\frac{i}{2}\left( \epsilon\partial_{t}\phi^{*}-\eta^{i}\partial_{i}\phi^{*}+v^{i}\left( t\partial_{i}\phi^{*}+imx_i\phi^{*}\right)\right) \partial_{t}\phi
\label{7}
\end{equation}
\begin{equation}
\frac{i}{2}\phi^{*}\delta_{0}(\partial_{t}\phi)=\frac{i}{2}\phi^{*}\left( \epsilon\partial_{t}(\partial_{t}\phi)-\eta^{i}\partial_{i}(\partial_{t}\phi)+\left(v^{i} t\partial_{i}-imv^ix_i\right)\partial_{t}\phi+v^{i}\partial_{i}\phi\right)
\label{8}  
\end{equation}
\begin{equation}
\frac{i}{m}\delta_{0}(\partial_{k}\phi^{*})\partial_{k}\phi=\frac{i}{m}\left[ \epsilon\partial_{t}(\partial_{k}\phi^{*})-\eta^{i}\partial_{i}(\partial_{k}\phi^{*})+\left(v^{i} t\partial_{i}+ imv^ix_i\right)\partial_{k}\phi^{*}+\partial_{k}\eta^{i}\partial_{i}\phi^{*}+ imv^{k}\phi^{*}\right] \partial_{k}\phi
\label{9}
\end{equation}
Considering only the contribution of the boost part of the global Galilean transformation, and using (\ref{c}), $\delta_{0}\mathcal{L}$ becomes,
\begin{eqnarray*}
&\delta_{0}\mathcal{L}=\frac{i}{2} [v^{i} t\partial_{i}\phi^{*}\partial_{t}\phi+ imv^{i}x_i\phi^{*}\partial_{t}\phi +\phi^{*}v^{i} t\partial_{i}\partial_{t}\phi-imv^{i}x_{i}\phi^{*}\partial_{t}\phi+\phi^{*}v^{i}\partial_{i}\phi\\&-v^{i}t \partial_{i}\phi\partial_{t}\phi^{*}+imx_{i}v^{i}\phi \partial_{t}\phi^{*}-\phi v^{i} t\partial_{i}\partial_{t}\phi^{*}-imv^{i}x_{i}\phi \partial_{t}\phi^{*}-\phi v^{i}\partial_{i}\phi^{*}]\\&-\frac{1}{2m}\left[ \left(v^{i} t\partial_{i}+ imv^ix_i\right)\partial_{k}\phi^{*}+ imv_{k}\phi^{*}\right] \partial_{k}\phi\\&-\frac{1}{2m}\left[ \left(v^{i} t\partial_{i}-imv^ix_i\right)\partial_{k}\phi-imv^{k}\phi\right] \partial_{k}\phi^{*}\\
\end{eqnarray*}
which reduces to,
\begin{equation}
\delta_{0}\mathcal{L}=v^{i}t\partial_{i}\mathcal{L}
\label{12}
\end{equation}
We are therefore led to the following result,
\begin{align}
\triangle \mathcal{L} &=\delta_{0}\mathcal{L}+{\xi}^{\mu}{\partial}_{\mu}\mathcal{L}\nonumber\\ 
&= v^{i}t {\partial}_{i}\mathcal{L}-v^{i}t\partial_{i}\mathcal{L}\nonumber\\ & =0
 \label{11}
\end{align}
In a similar manner, it can be shown that the other parts of the global Galilean transformation also render the action invariant. We thus conclude that the action is invariant under the global Galilean transformation
 
 Note that in addition to the global Galilean symmetry, the theory (\ref{globalaction}) is invariant under the global phase transformation, $\delta_0 \phi=i\alpha\phi$ where $\alpha$ is the gauge transformation parameter. 

  To obtain the corresponding theory of (\ref{globalaction}) invariant under the 3-d diffeomorphism (\ref{3ddiff}) we follow the prescription of (\ref{localaction1}). This leads to the action 
\begin{equation}
S = \int dt  \int d^3x \sqrt{g}\left[ \frac{i}{2}\left( \phi^{*}\nabla_{t}\phi-\phi\nabla_t\phi^{*}\right) -\frac{1}{2m}\nabla_a\phi^{*}\nabla_a\phi\right]
\label{localschrodinger} 
\end{equation}
Now
\begin{align}
\nabla_a\phi^{*}\nabla_a\phi &=\delta^{ab}\nabla_a\phi^{*}\nabla_b\phi\notag\\&=
\delta^{ab}{\Sigma_a}^k{\Sigma_b}^l D_k\phi^{*}D_l\phi\notag\\&=g^{kl}D_k\phi^{*}
D_l\phi\label{nkl}
\end{align}
where
\begin{equation}
g^{kl} = \delta^{ab}{\Sigma_a}^k{\Sigma_b}^l
\label{metricin}
\end{equation}
Using (\ref{metric1}) and (\ref{sl}) we observe that
\begin{equation}
g^{kl}g_{ln} = \delta^{k}_n \label{gklin}
\end{equation}
Hence $g^{kl}$ is indeed the inverse metric.

Note the dual role played by the field ${\Sigma_a}^k$. In section 2 they were introduced to define the local covariant derivatives from the global covariant derivatives in a theory defined on flat space. Now in the above equation they act as vielbeins connecting the tangent space and the curved 3-d space on which the theory is now formulated. This dual aspect which is observed here has an analogy in Poincare gauge theory as has been discussed for instance in \cite{Blagojevic:2002du} (page 61). 

Using (\ref{nkl}) in (\ref{localschrodinger}) we obtain the most general 3-d diffeomorphism invariant Schrodinger action as,
\begin{equation}
S = \int dt  \int d^3x \sqrt{g} \left[ \frac{i}{2}\left( \phi^{*}\nabla_{t}\phi-\phi \nabla_t\phi^{*}\right) -\frac{1}{2m}g^{kl}D_k\phi^{*}D_l\phi\right]
\label{diffschrodinger} 
\end{equation}
Such theories have recently been used in theoretical condensed matter physics. Note that it is very easy to take the flat limit of (\ref{diffschrodinger}), one just has to replace $g^{kl}$ by $\delta^{kl}$, and substitute covariant derivatives by ordinary derivatives. This immediately reproduces (\ref{globalaction}).

     As has been mentioned already in the introduction, the type of nonrelativistic diffeomorphism invariant models as (\ref{diffschrodinger}) are currently in vogue in condensed matter theory for the analysis of the motion of two dimensional trapped electrons \cite{SW}, specifically in connection with the study of FQHE. To understand this connection more explicitly let us consider a special case of (\ref{3ddiff}) where $\xi^i$ are time independent. From the second equation of the set(\ref{delth}) we find that $\Psi^k=0$ is admissible. 
 Substituting these in (\ref{diffschrodinger}) we get 
\begin{equation}
S = \int dt  \int d^3x {\sqrt{g}} \left[ \frac{i}{2}\left( \phi^{*}\left(\partial_{t} + i A_t\right)\phi-\phi\left(\partial_{t} - i A_t\right)\phi^{*}\right)  -\frac{1}{2m}g^{kl}\left(\partial_{k} - i A_k\right)\phi^{*}\left(\partial_{l} + i A_l\right)\phi\right]
\label{sw} 
\end{equation}
Note that this action is exactly of the same form as diff. invariant nonrelativistic 3-d Schrodinger theory introduced in \cite{SW, H} as an effective field theory. In this limit we get from, (\ref{delA})
\begin{eqnarray}
&{\delta}_0 A_{k} = -{\partial}_k {\xi}^i A_i - {\xi}^i {\partial}_iA_k\nonumber\\
&\delta_0A_t=-\xi^i\partial_iA_t
\label{delA1}
\end{eqnarray}
These transformation relations also exactly match with those given in \cite{SW}. Thus our formulation systematically leads to the model of \cite{SW} in the static transformation limit.


  The question of time varying diffeomorphism parameters demands careful considerations. Such situations will automatically appear if we include local Galilean boost in the effective 3-d diffeomorhism parameter $\xi$.
The condition (\ref{delth}) clearly shows that when $\xi$ is time dependent, $\Psi_k$ cannot be zero. Thus one has to consider the general form (\ref{diffschrodinger}) as the effective 3-d diffeomorhism invariant theory if one adopts the approach of constructing the nonrelativistic diffeomorphism invariant model by systematic localisation of Galilean symmetry. However there could be some other approach leading to a different nonrelativistic diffeomorphism invariant model. An example of this sort is provided by Hoyos and Son \cite{H} where a different model has been constructed which has diffeomorphism under spatial translation. Moreover we have seen here that if 
we adopt the restriction of no time translation and further consider only time independent transformations then our results reproduce those of \cite{H}.

\section{Conclusions}
We have discussed a method of localisation of the global Galilean invariance of a general field theoretic model. Local coordinate systems were erected at the different points of 3-d space. Though the connections between these local bases with the global basis are trivial, the distinction has been proved to be very useful in constructing the covariant derivatives. We have first defined covariant derivatives with respect to the global coordinates and then transformed them to covariant derivatives with respect to the local coordinates. New fields are introduced in the process, the transformation of which are determined so that the local covariant derivatives transform under local Galilean
transformation as the ordinary derivatives under global Galilean transformations. The localisation of the transformations also implies a change in the measure of integration. We have shown that the measure can be corrected appropriately by some functions of the newly introduced fields that were involved in the transformation of the global to local covariant derivatives. Substitution of the ordinary derivatives by the local covariant derivatives in the original global Galilean invariant action and correcting the measure of integration appropriately we obtain an action invariant under the local Galilean transformations. This procedure, as one can see, follows the Utiyama method of the localising the Poincare gauge transformation \cite{Utiyama:1956sy, Kibble:1961ba, sciama}. \footnote{The Utiyama method has been used in the context of nonrelativistic particle model \cite{PLP, Stichel, LSZ}, but as far as we know, it has thus far not been applied in the context of Galilean invariant field theories. }

   The process of localisation of the global Galilean invariance of field theories has been shown to provide a systematic algorithm
for the construction of 3-d diffeomorphism invariant field theories which have been used ubiquitously in the theory of Fractional quantum Hall effect(FQHE) \cite{SW, H} in recent times. Our approach may be contrasted with the way of introducing such models as effective field theories. 
In the gauging of Galilean invariance several new fields are introduced. These new fields along with their transformations obtained here will definitely be useful in phenomenological model building in theoretical condensed matter physics.

 The connection of our formulation with the research in condensed matter physics is however only an important special application of the formalism developed in this letter. Comparison with the Poincare gauge theory suggests that the process of localisation of the Galilean invariance should have connection with the geometric Newton Cartan theory of Newtonian gravity. In fact in the special case where time translation is ignored, our theory is already equivalent to matter coupled with external gravitational field. The next step is to give dynamics of the new fields introduced in the localisation procedure. One may guess that the whole procedure can be given a geometric parallel, and  thus we should be able to arrive at the Newton Cartan geometry as a fall out of the localisation.
These and similar works are currently in progress.

\end{document}